\documentclass[12pt,preprint]{aastex}
\newcommand{\siml}{\lower4pt \hbox{$\buildrel < \over \sim$}}
\newcommand{\simg}{\lower4pt \hbox{$\buildrel > \over \sim$}}
\def\paren#1{\left( #1 \right)}

\received{2003 April 3}
\begin{document}
\title{Early Optical Afterglows from Wind-Type Gamma-Ray Bursts}
\author{Shiho Kobayashi$^{1,2}$ and  Bing Zhang$^1$}
\affil{ $^{1}$Department of Astronomy \& Astrophysics, Pennsylvania State
University, University Park, PA 16802\\
$^{2}$Center for Gravitational Wave Physics, Pennsylvania State University,
University Park, PA 16802}
\begin{abstract}
We study prompt optical emission from reverse shocks in the wind-type
gamma-ray bursts. The emission is evaluated in both the thick and
thin shell regimes. We discuss the angular time delay effect and the
post-shock evolution of the fireball ejecta, which determine the decay
index of the prompt optical emission and the duration of the radio
flare. We discuss distinct emission signatures of the wind environment
compared with the constant interstellar medium environment. We also
present two recipes for directly constraining the initial 
Lorentz factor of the fireball using the reverse and forward
shock optical afterglow data for the wind case.
\end{abstract}


\section{Introduction}

The gamma-ray burst (GRB) afterglow observations are usually explained
by a model in which a relativistic fireball shell (ejecta) is
expanding into a uniform interstellar medium (ISM). However, there is
observational evidence suggesting a link between GRBs and massive
stars or star formation (e.g. M\'esz\'aros 2001 for a review). An
important consequence of a massive star origin for the afterglow is
that the fireball shell should be expanding into a pre-burst stellar
wind of the progenitor star with a density distribution of $\rho
\propto R^{-2}$ (e.g. Chevalier \& Li 1999; M\'esz\'aros, Rees \&
Wijers 1998; Dai \& Lu 1998). This wind model is discussed to be
consistent with some GRB afterglow data (Chevalier \& Li 1999, 2000).

It is expected that many early optical afterglows will be discovered
soon in the observational campaigns led by HETE-2 and Swift. Long time
span (from right after the GRB trigger to a year) observational data
will allow us to distinguish differences between the ISM and wind
models clearly.

In this paper, we will discuss the optical reverse shock emission for
the wind model in detail. The previous study (Chevalier \& Li 2000)
gave the discussions for the thick shell case, and we will consider
both the thin and thick shell cases. We show that the angular time
delay effect plays an important role in discussing the light curve
decaying phase of the reverse shock emission.

\section{The Model}

We consider a relativistic shell (fireball ejecta) with an isotropic
energy $E$, an initial Lorentz factor $\eta$ and an initial width
$\Delta_0$ expanding into a surrounding medium with density distribution
of $\rho=A R^{-2}$ (wind model). In this paper, Lorentz factors
$\gamma$, radii $R$ and widths $\Delta$ are measured in the laboratory
frame where the progenitor is at rest. Thermodynamic
quantities: mass densities $\rho$ and internal energy $e$ are measured
in the fluid comoving frame.
The interaction between the shell and the wind is described by two
shocks: a forward shock propagating into the wind and a reverse shock
propagating into the shell. The shocks accelerate electrons in the
shell and wind material, and the electrons emit photons via
synchrotron-cyclotron process.  For the synchrotron process, the
spectrum of each shock emission 
is described by  a broken power law with
a peak $F_{\nu, max} \propto N_e B \gamma$ and break frequencies:
a typical frequency $\nu_{m} \propto B\gamma \gamma_m^2$ and
a cooling frequency $\nu_{c}\propto 1/B^3\gamma t^2$ (Sari, Piran \&
Narayan 1998) where  $N_e$ is the number of electrons accelerated by the
shock, $B$ is the magnetic field strength behind the shock, $\gamma$ and
$\gamma_m$ are the bulk Lorentz factor of the shocked material and
the random Lorentz factor of the typical electrons in the shocked
material, respectively. Assuming that constant fractions ($\epsilon_e$
and $\epsilon_B$) of the internal energy $e$ produced by the shock
go into the electrons and the magnetic field, we get
$\nu_m \propto \gamma \rho^{-2} e^{5/2}, 
\nu_c \propto \gamma^{-1}e^{-3/2} t^{-2}$ and
$F_{\nu,max}\propto N_e\gamma e^{1/2}$.

\section{Forward Shock}

Observations of optical afterglows usually start around several hours
after the burst trigger. At such a late time, the shocked wind material
forms a relativistic blast wave and carries almost all the energy of the
system. Chevalier \& Li (1999) gave the  characteristics of
the forward shock (blast wave) emission as follows,
\begin{eqnarray}
\nu_{m,f}(t) &\sim&  1.6 \times 10^{12} \zeta^{1/2}
\epsilon_{e,-1}^2 \epsilon_{B,-2}^{1/2}
E_{52}^{1/2}t_{d}^{-3/2} ~\mbox{Hz},
\label{eq:nmf}
\\
\nu_{c,f}(t) &\sim& 6.3 \times 10^{13} \zeta^{-3/2}
\epsilon_{B,-2}^{-3/2}
E_{52}^{1/2}A_\ast^{-2}t_{d}^{1/2} ~\mbox{Hz},
\label{eq:ncf}
\\
F_{\nu,max,f}(t) &\sim& 4.5 ~d^{-2}
\zeta^{3/2}\epsilon_{B,-2}^{1/2}
E_{52}^{1/2}A_\ast t_{d}^{-1/2} ~\mbox{mJy},
\label{eq:fnmax}
\end{eqnarray}
where $\zeta=(1+z)/2$, $d=d_L(z)/(2\times10^{28}$cm), $z$ and $d_L(z)$ 
are the redshift and luminosity distance of the burst, respectively,  
$d_L(1)\sim 2\times 10^{28}$cm for the standard cosmological
parameters ($\Omega_m=0.3$, $\Omega_\Lambda=0.7$ and $h=0.7$),
$\epsilon_{e,-1}=\epsilon_e/0.1$, 
$\epsilon_{B,-2}=\epsilon_B/0.01$,
$E_{52}=E/10^{52}$ ergs, $A_\ast=A/5\times10^{11}$ g
cm$^{-1}$, $t_d$ is the observer's time in units of days.
The optical flux from the forward shock decays proportional to
$t^{-1/4}$ initially, and decays faster as $t^{-(3p-2)/4}$ after
a transition of  $\nu_{m,f}$ through the optical band
$\nu_R \sim 5\times 10^{14}$Hz (Chevalier \& Li 2000)
where  $p$ is the index of the power law distribution of the
accelerated electrons.
Using $\nu_{R\ast}=\nu_R/5\times10^{14}$Hz, the break time $t_{m,f}$
(the passage of $\nu_{m,f}$) and the optical flux (in the fast-cooling
regime) at that time are
\begin{eqnarray}
t_{m,f} &\sim& 30 ~\zeta^{1/3} \epsilon_{e,-1}^{4/3} \epsilon_{B,-2}^{1/3}
E_{52}^{1/3} \nu_{R\ast}^{-2/3}~\mbox{min},
\label{eq:tmf}\\
F_{\nu_R,f}(t_{m,f}) &\sim& 4 ~d^{-2} \zeta^{2/3} \epsilon_{e,-1}^{-1/3}
\epsilon_{B,-2}^{-1/3} E_{52}^{2/3} \nu_{R\ast}^{-1/3} ~\mbox{mJy}.
\label{eq:fnurf}
\end{eqnarray}

\section{Reverse Shock}

At earlier time when the reverse shock crosses the shell, the
forward-shocked wind and the reverse shocked shell carry comparable
amount of energy. A significant emission is expected from the reverse
shock also (M\'esz\'aros \& Rees 1997; Sari \& Piran 1999a).
During the reverse shock crossing, there are four regions
separated by the two shocks: the wind (denoted by the subscript 1), the
shocked wind (2), the  shocked shell material (3) and the unshocked shell
material (4). Using the jump conditions for the shocks and the equality
of pressure and velocity along the contact discontinuity, we can
estimate the Lorentz factor $\gamma_i$, the internal energy $e_i$ and the mass
density $\rho_i$ in the shocked regions as functions of three variables
$\gamma_4 (=\eta)$, $\rho_1$ and  $\rho_4$ (Sari and Piran 1995).

There are two limits to get a simple analytic solution to the
hydrodynamic quantities at a shell radius $R$ (Sari and Piran 1995). If
the Lorentz factor is low $\eta^2 \ll f$ where $f=\rho_4/\rho_1$, the
reverse shock is Newtonian which means that the Lorentz factor of the
shocked shell material is almost unity in the frame of the unshocked
shell material. It is too weak to slow down the shell effectively so
that $\gamma_3\sim\eta$.  On the other hand, if the Lorentz factor is
high $\eta^2 \gg f$, the reverse shock is relativistic, and
considerably decelerates the shell material, hence
$\gamma_3 \sim \eta^{1/2} f^{1/4}$.

\subsection{Critical Radii}
Since the density ratio $f$ is generally a function of $R$, there is
a possibility that the reverse shock evolves from Newtonian to
relativistic during the propagation. The evolution of the reverse shock
depends on the ratio of two radii: $R_\gamma \equiv E/4\pi Ac^2 \eta^2$
where the forward shock sweeps a mass of $E/c^2\eta^2$ and   $R_s \equiv
\Delta_0 \eta^2$ where  the shell begins to spread if the initial
Lorentz factor varies by order $\eta$ (Sari \& Piran 1995; Kobayashi,
Piran \& Sari 1999). Another important radius is $R_\times$ where the
reverse shock crosses  the shell. The lab-frame time it takes for the
reverse shock to cross a width $dx$ of the shell material is given by
$dt_{lab} \sim dR/c \sim \eta f^{1/2} dx/c$
(Sari \& Piran 1995).
We can regard $R/c$ as time $t_{lab}$ in the laboratory frame
because of the highly relativistic expansion of the shell. Since the
whole shell width is $\Delta\sim\max[\Delta_0, R/\eta^2]$, we obtain
$R_\times \sim \max[ (R_sR_\gamma)^{1/2}, R_\gamma]$.

Considering $\eta^2/f=\max[R_s/R_\gamma, R/R_\gamma]$, we
can classify the evolution of reverse shocks into two cases by using
a critical Lorentz factor $\eta_c \equiv (E/4\pi Ac^2\Delta_0)^{1/4}$
(see Sari \& Piran 1995 and Kobayashi \& Zhang 2003 for the ISM model).
If $R_s/R_\gamma=(\eta/\eta_c)^4 >1$ (Thick Shell Case), the reverse
shock is relativistic from the beginning (at the end of the GRB
phase), which is different from the ISM model in which the reverse
shock only becomes relativistic later. The reverse shock crosses the
shell at $R_\times \sim (R_sR_\gamma)^{1/2}$ before the shell begins to
spread at $R_s$,  and it significantly decelerates the shell material
$\gamma_3 \sim \eta_c$. If $R_s/R_\gamma < 1$ (Thin Shell Case), the
reverse shock is initially Newtonian  and becomes only mildly
relativistic when it traverses the shell at  $R_\times\sim R_\gamma$. We
can regard $\gamma_3$ as constant ($\sim \eta$) during the shock crossing.

\subsection{Synchrotron Emission}

Since the wind density at the initial interaction is much larger than
the medium density for the ISM model, the cooling frequency $\nu_{c,r}$ 
of the reverse shock emission
is much lower than the typical (injection) frequency $\nu_{m,r}$ in the 
wind model. The random  Lorentz factor $\gamma_c$ of
the electrons that radiate at the cooling frequency $\nu_{c,r}$ could be
sub-relativistic or Newtonian, $\gamma_c (t_\times) \sim  1
\zeta^{-7/4}\epsilon_{B,-2}^{-1} A_\ast^{-5/4}E_{52}^{1/4}
t_{\times,\ast}^{3/4}$ for our typical parameters.  
This makes the radiation mechanism cyclotron radiation at low frequencies 
$\sim \nu_{c,r}$ and at early times $< t_\times$. However, the detailed 
modelling on the cyclotron emission is not important, because the flux 
is suppressed and determined by the self absorption. (We will discuss the
self-absorption at the end of this subsection.) On the other hand, the random 
Lorentz factor $\gamma_{\nu_R}$ of the electrons corresponding to the optical
frequency $\nu_R$ is relativistic. Since the electron distribution
$N(\gamma_e)$ around $\gamma_{\nu_R}$ (and above it) is determined by
the distribution of injected electrons at the shock, which are
relativistic, and by the synchrotron radiation cooling, we apply the
conventional synchrotron  model to estimates the light curve of optical
flashes. 

The observer time $t\equiv (1+z)R/c\gamma^2$ is proportional to $R$,
because the Lorentz factor of the shocked shell during
the shock crossing $\gamma_\times \sim \min[\eta, \eta_c]$
is constant. By using the shock jump conditions, one finds that the
scalings before the crossing time $t_\times = (1+z)
R_\times/c\gamma_\times^2$ are given  
by  $e_3 \propto t^{-2}$, $\rho_3 \propto t^{-2}$ and
$N_e\propto  t$    in the thick shell case, and 
$e_3 \propto t^{-2}$, $\rho_3 \propto t^{-3}$ and
$N_e\propto  t^{1/2}$  in the thin shell case.
The scalings of the spectral characteristics  at $t<t_\times$ are
$\nu_{m,r} \propto t^{-1},  \nu_{c,r} \propto t$ and
$F_{\nu,max,r}\propto t^0$ for thick shell and 
$\nu_{m,r} \propto \nu_{c,r} \propto t $ and 
$F_{\nu,max,r} \propto t^{-1/2}$ for thin shell. The scalings of
$\nu_{c,r}$ and $F_{\nu,max,r}$ themselves are not correct, but the
optical flux estimated by these scalings are right. We evaluated
the scalings to calculate the optical light curve.

The initial shell width $\Delta_0$ is given by the intrinsic
duration of the GRB, $\Delta_0 \sim (1+z)^{-1} cT$ (Kobayashi, Piran \&
Sari 1997), the shock crossing time $t_\times$ can be
written in the following form,
\begin{equation}
t_\times \sim \paren{\frac{\gamma_\times}{\eta_c}}^{-4}T,
\label{eq:ttimes}
\end{equation}
where $\eta_c \sim 60 ~\zeta^{1/4} E_{52}^{1/4} A_\ast^{-1/4}
T_1^{-1/4}$ and $T_1=T/10$sec. We can determine the critical Lorentz 
factor $\eta_c$ from the observations of the GRB and afterglow. If we
detect the shock crossing time $t_\times$ (the reverse shock peak time),
the Lorentz factor during the shock crossing time $\gamma_\times$
can be estimated from eq. (\ref{eq:ttimes}).

The spectral characteristics of the reverse shock emission
at $t_\times$ are related to those of the forward shock emission by the
following simple formulae 
\footnote{ In this paper, we assume that $\epsilon_B$, $\epsilon_e$ 
and $p$ are the same for two shocked regions. Some recent works
(Zhang, Kobayashi \& M\'esz\'aros 2003; Kumar \& Panaitescu 2003; 
Coburn \& Boggs 2003) show that $\epsilon_{B,r}$ might be larger
than $\epsilon_{B,f}$ where the subscripts 'r' and 'f' indicate reverse
and forward shocks, respectively. In such a case, the formulae are
replaced by  eqs (3)-(5) in Zhang et al. (2003). Though the 
power $F_{\nu,max,r}$ increases by a factor of 
$(\epsilon_{B,r}/\epsilon_{B,f})^{1/2}$ compared to a case of 
$\epsilon_{B,r}=\epsilon_{B,f}$, the cooling frequency $\nu_c$ 
decreases by a factor of  $(\epsilon_{B,r}/\epsilon_{B,f})^{3/2}$.
This results in a dimmer optical reverse emission at the peak time 
by factor of $(\epsilon_{B,r}/\epsilon_{B,f})^{-1/4}$. Note that 
in the ISM model, the reverse shock emission is in slow cooling regime
and that $\epsilon_{B,r} > \epsilon_{B,f}$ gives a brighter
reverse shock emission (see Zhang et al. 2003).
}
(Kobayashi \&Zhang 2003),
\begin{equation}
\nu_{m,r}(t_\times) \sim \frac{\eta^2}{\gamma_\times^4}
\nu_{m,f}(t_\times),  \ \ \
\nu_{c,r}(t_\times) \sim \nu_{c,f}(t_\times), \ \ \
F_{\nu,max,r} (t_\times) \sim
\frac{\gamma_\times^2}{\eta} F_{\nu,max,f} (t_\times).
\label{eq:relation}
\end{equation}
Using eqs. (\ref{eq:nmf}), (\ref{eq:ncf}) and (\ref{eq:fnmax}), we get
\begin{eqnarray}
\nu_{m,r}(t_\times) &\sim& 
5.0\times10^{14} ~\zeta^{-1/2} \epsilon_{e,-1}^2 \epsilon_{B,-2}^{1/2}
E_{52}^{-1/2}A_\ast \eta_2^2 t_{\times,\ast}^{-1/2} ~\mbox{Hz}
\label{eq:numr} \\
\nu_{c,r}(t_\times) &\sim&  
1.5\times10^{12} ~\zeta^{-3/2} \epsilon_{B,-2}^{-3/2} 
E_{52}^{1/2}A_\ast^{-2} t_{\times,\ast}^{1/2}~\mbox{Hz}
\label{eq:nucr} \\
F_{\nu,max,r}(t_\times) &\sim& 
3.0 ~d^{-2} \zeta^2 \epsilon_{B,-2}^{1/2}
E_{52}A_\ast^{1/2} \eta_2^{-1} t_{\times,\ast}^{-1}  ~\mbox{Jy}
\label{eq:Fr}
\end{eqnarray}
where $\eta_2=\eta/100$ and $t_{\times,\ast}=t_\times/50$sec.
The reverse shock emission is in the fast
cooling regime $\nu_{c,r} < \nu_{m,r}$ during the shock crossing.
So for $t<t_\times$, the optical flux from the reverse shock increases
as $\propto t^{1/2}$ for thick shell case ($\nu_R<\nu_{m,r}(t_\times)$) 
(Chevalier \& Li 2000), and $\propto t^{(p-1)/2}$ for thin shell case 
($\nu_R > \nu_{m,r}(t_\times)$). Since the Lorentz
factor of the forward-shocked material is also constant during shock
crossing, the optical emission from the forward shock evolves as
$t^{1/2}$ at $t<t_\times$. But this component is usually masked by the
reverse shock emission.

At the shock crossing time $t_\times$, the optical flux reaches the
peak $F_{\nu_R,r}$
$\sim F_{\nu,max,r}$ $(\nu_R/\nu_{c,r})^{-1/2}$.  
\begin{equation}
F_{\nu_R,r}(t_\times) \sim 
160 ~d^{-2} \zeta^{5/4} \epsilon_{B,-2}^{-1/4} E_{52}^{5/4}
A_\ast^{-1/2} \nu_{R\ast}^{-1/2} \eta_2^{-1} t_{\times,\ast}^{-3/4} 
~\mbox{mJy}
\label{eq:opticalflash}
\end{equation}

Generally, using the time dependences of the spectral characteristics
as well as eq. (\ref{eq:relation}), we can
relate this peak flux with the optical flux of the forward shock at
the break time, 
\begin{equation}
F_{\nu_R,r}(t_\times)\sim \paren{\frac{\gamma_\times^2}{\eta}}^{2-a}
\paren{\frac{t_\times}{t_{m,f}}}^{(2-3a)/4}F_{\nu_R,f}(t_{m,f}).
\label{eq:peakbreak}
\end{equation}
where $a=p$ if $\nu_{m,r}(t_\times)$ is below the optical band, and $a=1$
if it is above.  Detection of the peak
$(t_\times, F_{\nu_R,r}(t_{\times}))$ and the break
$(t_{m,f}, F_{\nu_R,f}(t_{m,f}))$ will give a constraint on
the initial Lorentz factor $\eta$. A similar recipe has been
proposed by Zhang et al. (2003) for the ISM case.

Synchrotron self-absorption would reduce our estimate
(\ref{eq:opticalflash}) of the optical flash if it is optically thick.   
A simple way to account for this effect is to estimate the maximal flux
emitted by the shocked shell material as a blackbody (Sari \& Piran
1999b; Chevalier \& Li 2000). The blackbody flux at the optical band  
$\nu_R$ and at the peak time $t_\times$ is given by 
$F_{\nu_R,BB}\sim 16 ~\zeta^{5/4}d^{-2}\nu_{R\ast}^{5/2}$  
$\epsilon_{B,-2}^{-1/4}A_\ast^{-1}E_{52}^{3/4}$
$t_{\times,\ast}^{7/4}$ Jy. The self-absorption frequency at the
peak time is $\nu_{sa}\sim 1.0 \times 10^{14} 
A_\ast^{1/6}E_{52}^{1/6} \eta_2^{-1/3}t_{\times,\ast}^{-5/6}$ Hz.
Considering that the peak time $t_\times$ is larger than the burst
duration $T$,  the synchrotron self absorption will not affect
the peak flux of an optical flash significantly for long bursts with  
$T > 8 ~\nu_{R\ast}^{-6/5} A_\ast^{1/5}E_{52}^{1/5} \eta_2^{-2/5}$ sec. 
However, the self-absorption is important at early times ($<t_\times$) 
in the wind model.  The optical light curve should initially behave as
$t^{5/2}$  right after the GRB trigger because of the absorption and it
turns into $\sim t^{1/2}$ later  (see Fig 1). The break time is 
$t\sim 5 ~\nu_{R\ast}^{-3/2} A_\ast^{1/4}E_{52}^{1/4} \eta_2^{-1/2}
t_{\times,\ast}^{-1/4}$ sec.

\subsection{Angular Time Delay Effect}
The cooling frequency $\nu_{c,r}(t_\times)$ is well below the
optical band for our typical parameters, the optical emission from
the reverse shock should ``vanish'' after the peak. However, the
angular time delay effect prevents abrupt disappearance.
The optical light curve at $t>t_\times$ is determined by off-axis
emissions.  In the local frame, the spectral power is described by
a broken power law with a low and high frequency indices $-1/2$ and
$-p/2$ and the break frequency 
$\sim \nu_{m,r}(t_\times)/\gamma_\times$.
As we see higher latitude emissions, the blue shift effect due to
the relativistic expansion becomes smaller. The blue shifted break 
frequency passes through the optical band $\nu_R$ at time 
$\sim t_\times \nu_{m,r}(t_\times)/\nu_R$. The optical flux
initially evolves as $t^{-5/2}$ and decays as $t^{-(p+4)/2}$ after the
passage (Kumar \& Panaitescu 2000b). 

Since the forward shock emission decays slower $t^{-1/4}$, it
begins to dominate the optical band at
$t_{trans} \sim (t_\times^{4\alpha}/t_{m,f})^{1/(4\alpha-1)}
[F_{\nu_R,r}(t_\times)/F_{\nu_R,f}(t_{m,f})]^{4/(4\alpha-1)}$
where we assumed that the optical reverse shock emission decays
proportional to $t^{-\alpha}$. For our typical parameters,
$\nu_{m,r}(t_\times)$ comes to the optical band. Assuming
$\alpha = 13/4$ (p=2.5) (see fig 1(a) in which
$\alpha \sim 3$ is applied), we get
$t_{trans}/t_\times \sim 
2.5 ~\zeta^{1/6} E_{52}^{1/6} A_\ast^{-1/6} \eta_2^{-1/3}
t_{\times,\ast}^{-1/6}$. The optical emission from the reverse shock
drops below that from the forward shock within a time scale several
times of the peak time $t_\times$.

\subsection{Duration}

If the fireball ejecta is collimated in a jet with an opening angle
$\theta_j$, the duration of the reverse shock emission is
$t_{ang}\sim (1+z) \theta_j^2 R_\times/c$. The angle $\theta_j$ might be
determined from a jet break time  $t_j\sim (\theta_j
\gamma_\times)^4t_\times$ of the forward shock
emission (Rhoads 1999; Sari,Piran \& Halpern 1999), even though  a jet
break in the wind model may not be as clear as that in the ISM model 
(Kumar \& Panaitescu 2000a; Gou et al. 2001).
\begin{equation}
t_{ang} \sim \paren{t_\times t_j}^{1/2} \sim
35 ~ \paren{\frac{t_\times}{\mbox{50 sec}}}^{1/2}
\paren{\frac{t_j}{\mbox{1 day}}}^{1/2} ~\mbox{min}
\end{equation}

At a low frequency $\nu < \nu_{c,r}(t_\times)$, the observer receives
photons from the fluid element on the line of the sight until a
break frequency $\nu_{cut}$, which is equal to $\nu_{c,r}(t_\times)$ at
$t_\times$, crosses the observational band. We now consider this time scale.
After the reverse shock crosses the shell, the profile of the
forward-shocked wind begins to approach the Blandford-McKee (BM)
solution (Blandford \& McKee 1976; Kobayashi, Piran \& Sari 1999). Since
the shocked shell is located not too far behind the forward shock, it
roughly fits the BM solution.  A given fluid element in a blast wave
evolves as $\gamma \propto t^{-3/8}$, $\rho \propto t^{-9/8}$ and
$e\propto t^{-3/2}$. Assuming that the electron energy and
the magnetic field energy remains constant fractions of the internal energy
density of the shocked shell, the emission frequency of each electron
with $\gamma_e$ drops quickly with time according to  
$\nu_e \propto B \gamma \gamma_e^2 \propto t^{-15/8}$ ($\gamma_e \gg 1$)
or $\nu_e \propto B \gamma \propto t^{-9/8}$ ($\gamma_e \sim 1$).
Using the scaling for the cyclotron emission,  $\nu_{cut}$ passes
through the observational frequency $\nu$ at  
\begin{equation}
t_{cut}\sim 70
\zeta^{-4/3} \epsilon_{B,-2}^{-4/3} E_{52}^{4/9}A_\ast^{-16/9} 
\nu_{10}^{-8/9} t_{\times,\ast}^{13/9} ~\mbox{min},
\label{eq:tcut}
\end{equation}
where $\nu_{10}=\nu/10$GHz. Though at the radio band and early times, self
absorption significantly reduce the flux,
$t_{dur}=\max[t_{ang},t_{cut}]$ can give a rough estimate of the
duration of the radio reverse shock emission.

\section{Case Studies}
GRB 990123: The optical flash (Akerlof et al. 1999) and radio flare
(Kulkarni 1999) associated with this burst are explained well by a
reverse shock emission in the ISM model 
(Sari \& Piran 1999b; Kobayashi \& Sari 2000).  
The basic parameters of this burst include (e.g. Kobayashi \& Sari 2000
and reference therein) $E_{52}\sim 140$, $z\sim 1.6$, $T_1\sim
6.3$ and $t_\times \sim 50$ sec. The wind model predicts a flatter
rising of $t^{1/2}$, $t^{(p-1)/2}$ or $t^{5/2}$ compared to $t^{3.4}$ evaluated
from the first two ROTSE data, and a steeper decline of $t^{-5/2}$ or 
$t^{-(p+4)/2}$ compared to the observations after the peak  $t^{-2}$ 
(Chevalier \& Li 2000). The optical reverse shock emission is expected
to be overtaken by that from the forward shock at  
$t_{trans} \sim 5 A_\ast^{-1/6} \eta_2^{-1/3}$  min, but the observed flash
decays as a single power law of $t^{-2}$ until it falls off below the
detection threshold at $\sim 11$ min. Using  the jet break time
$t_j \sim 2$ day (Kulkarni et al 1999), we obtain  $t_{ang} \sim 50$ min  
and $t_{cut}\sim 7.4 \epsilon_{B,-2}^{-4/3} A_\ast^{-16/9}$ hr. 
The reverse shock emission should disappear well before the radio flare
at $\sim 1$ day. We conclude that the wind model is inconsistent with
the observations. 

GRB 021004:  In the ISM model, it was shown that the major bump observed
in the afterglow light curve around $\sim 0.1$ day after the burst could
be explained by the passage of the typical frequency of the forward
shock emission through the optical band, and that 
the early time optical emission is a combination of reverse and forward
shock emissions (Kobayashi \& Zhang 2003). In the wind model, the optical
light curve of the forward shock emission initially 
behaves as $t^{-1/4}$ and decays steeper as $t^{-(3p-2)/4}$
after the typical frequency crosses the optical band.
The bump might be explained by the passage of $\nu_{m,f}$
\footnote{After the completion of our paper, we noticed that
Li and Chevalier (2003) gave a detailed studied on this possibility
in a recent paper.}. The basic parameters of this burst are
(e.g. Kobayashi \& Zhang 2002 and reference therein) 
$E_{52}\sim 5.6$, $z\sim 2.3$ and $T_1\sim 10$.
The critical Lorentz factor is $\eta_c \sim 60 A_\ast^{-1/4}$.
Assuming $t_{m,f}\sim 0.1$day  and $F_{\nu_R,f}(t_{m,f})\sim 1$mJy,
we obtain $\epsilon_e \sim 0.11$ and $\epsilon_{B}\sim 0.085$
from eqs (\ref{eq:tmf}) and (\ref{eq:fnurf}).  Since 
these are the typical values obtained in other afterglow observations
(Panaitescu \& Kumar 2002), the wind model might also fit the
observational data.

\section{Conclusions}

We have studied the optical emissions from reverse shocks
for the thin and thick shell cases. The differences between this model
and the ISM model are highlighted in Figure 1. In the ISM model, the
prompt optical emission increases proportional to $t^{5}$ for the the
thin shell case or $t^{1/2}$ for the thick shell case (Kobayashi
2000). In the wind model, it behaves as $t^{1/2}$ for the both
cases. The synchrotron self-absorption is important at early times
in this model. The luminosity increases as $t^{5/2}$ with a steep 
spectral index $F_{\nu}\propto \nu^{5/2}$ at the beginning. If a rapid
brightening with an index larger than
$5/2$ is caught, it could be an indication of the ISM-type GRB. 
The decay index of the emission is determined by the angular time delay
effect in the wind model so that $\sim t^{-3}$, while it depends on the
hydrodynamic evolution of the fireball ejecta in the ISM model hence 
$\sim t^{-2}$.

When we detect the peak time of the reverse shock emission, we can
estimate the Lorentz factor at that time from eq.(\ref{eq:ttimes}).  In
the ISM model, the peak time was given by a similar relation $t_\times
\sim (\gamma_\times/\eta_c^\prime)^{-8/3} T$ (Sari \& Piran 1999a) where
$\eta_c^\prime$ is a critical Lorentz factor and given by eq. (7)
in Kobayashi \& Zhang (2003). Additionally, if we detect the break,
caused by the passage of $\nu_{m,f}$, in the late time ($\sim 1$ hr
after the burst) optical light curve, from eq. (\ref{eq:peakbreak})
we can give another constraint on the initial Lorentz factor $\eta$
(see also Zhang et al. 2003 for the ISM case).

\acknowledgements This work is supported by NASA NAG5-9192 and the
Pennsylvania State University Center for Gravitational Wave Physics,
which is funded by NSF under cooperative agreement PHY 01-14375.


\noindent {\bf References}\newline
Akerlof,C.W. et al. 1999, Nature, 398, 400.\newline
Blandford,R.D. \& McKee,C.F. 1976, Phys of Fluids, 19, 1130.\newline
Chevalier,R.A. \& Li,Z.Y. 1999 ApJ, 520, L29.\newline
Chevalier,R.A. \& Li,Z.Y. 2000 ApJ, 536, 195.\newline
Coburn,W. \& Boggs,S.E. 2003, Nature, 423, 415.\newline
Dai,Z.G. \& Lu,T. 1998, MNRAS, 298, 87.\newline
Gou,L.J.,Dai,Z.G.,Huang,Y.F. \& Lu,T. 2001 A\&A, 368, 464.\newline
Li,Z.Y. \& Chevalier,R.A. 2003 ApJ, 589, L69.\newline
Kobayashi,S 2000, ApJ, 545, 807.\newline
Kobayashi,S, Piran,T. \& Sari,R. 1997, ApJ, 490, 92.\newline
Kobayashi,S, Piran,T. \& Sari,R. 1999, ApJ, 513, 669.\newline
Kobayashi,S \& Sari,R. 2000, ApJ, 542, 819.\newline
Kobayashi,S \& Zhang,B. 2003, ApJ, 582, L75.\newline
Kulkarni, S.R. et al. 1999, Nature, 398, 389.\newline
Kumar,P. \& Panaitescu,A. 2000a, ApJ, 541, L9.\newline
Kumar,P. \& Panaitescu,A. 2000b, ApJ, 541, L51.\newline
Kumar,P. \& Panaitescu,A. 2003, astro-ph/0305446.\newline
M\'esz\'aros,P. 2001, Science, 291, 79. \newline
M\'esz\'aros,P. \& Rees,M.J. 1997, ApJ, 476, 231.\newline
M\'esz\'aros,P., Rees,M.J. \& Wijers,R.A.M.J. 1998, ApJ, 499, 301.\newline
Panaitescu,A. \& Kumar,P. 2002, ApJ, 571, 779.\newline
Rhoads,J.E. 1999 ApJ, 525, 737.\newline
Sari,R. \& Piran,T. 1995 ApJ, 455, L143.\newline
Sari,R. \& Piran,T. 1999a ApJ, 520, 641.\newline
Sari,R. \& Piran,T. 1999b ApJ, 517, L109.\newline
Sari,R., Piran,T. \& Halpern,J.P. 1999 ApJ, 519, L17.\newline
Sari,R., Piran,T. \& Narayan,R. 1998 ApJ, 497, L17.\newline
Zhang,B., Kobayashi,S. \& M\'esz\'aros,P. 2003, ApJ, in press
(astro-ph/0302525).\newline

\clearpage
 \begin{figure}
\plotone{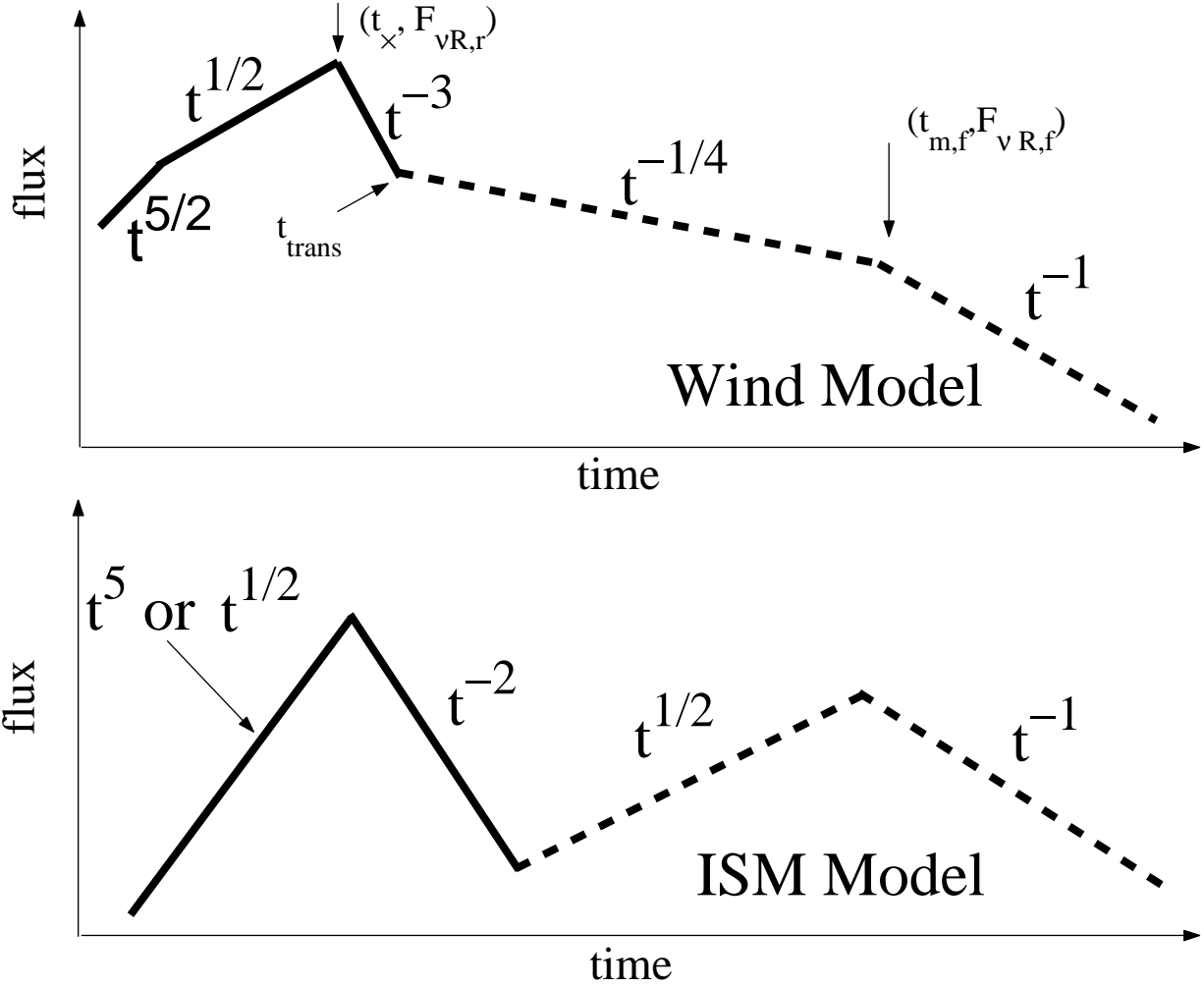}
\caption{Optical light curve: Wind Model and ISM Model.
Reverse Shock Emission (solid) and Forward Shock Emission
(dashed).
 \label{fig1}}
 \end{figure}
\end{document}